\documentclass[twocolumn]{jpsj3}
\usepackage{txfonts}

\title{Flow and Tilt-Induced Orientation of the Moving Vortex Lattice in Amorphous NbGe Superconducting Thin Films}

\author{Nobuhito KOKUBO$^1$\thanks{E-mail: kokubo@pc.uec.ac.jp},  Tetsuya YOSHIMURA$^2$, and Bunjyu SHINOZAKI$^2$}
\inst{$^1$Department of Engineering Science, The University
of Electro-Communication, Tokyo 182-8585, Japan \\
$^2$Department of Physics, Kyushu University, Fukuoka
812-8581, Japan } 

\abst{The orientation and deformation of moving vortex lattices in
the flux-flow state have been investigated in amorphous
superconducting NbGe thin films. Employing a mode-locking technique,
we detect how moving lattices deform and their orientation changes
as a magnetic field is tilted from normal to the film surface. For
high tilt angles the lattice orientation is aligned parallely with
the tilt direction. Meanwhile for low tilt angles the lattice
orientation depends on the vortex velocity and a velocity-induced
reorientation occurs. The characteristic velocity for the
reorientation varies remarkably as the moving lattices deform. The
observed features are consistent with an extended bond-fluctuation
theory, revealing that the anisotropic shaking vortex motion is
essential for determining the orientation of moving vortex
lattices.}

\kword{vortex lattice, reorientation, mode locking, flux flow,
amorphous superconductor}

\begin{document}
\maketitle

\section{Introduction}

Over the past few decades there has been considerable interest in
the orientation of a hexagonal vortex lattice (VL) in type II
superconductors \cite{Takanaka1971}. Within the London approximation
for isotropic superconductors all the possible orientations of the
vortex lattice are degenerate in
energy\cite{Campbell1988,KoganPRB1995}, while this is lifted not
simply by tilting an applied magnetic field \cite{BollePRL1993,
HessPRB1994}, but also when driven by applying a transport current.
In experiments various imaging techniques such as Bitter decoration
\cite{Pardonature}, small angled neutron scattering
\cite{Yaron,Eskildsen2011} and scanning tunneling microscopy
\cite{Troyanovski,KohenAPL,Kesprivate} have provided structural
evidence for the hexagonal order in moving VLs, and some of them
have shown that the lattice orientation (one of close-packed
directions of the hexagonal VL) is aligned to be either parallel or
perpendicular to the flow direction. These flow-induced
orientations, including a reorientation between them, have been
observed on both crystalline \cite{Pardonature,KohenAPL} and
non-crystalline weak pinning superconductors
\cite{kkbphysica2008,kkbphysica2010,OkumaPRB2009}, and thus
underlying crystallographic orientations are not at play. It is
rather essential to understand how quenched disorder affects the
orientation of moving VLs. Schmid and Hauger pointed out, for a long
time ago, that the dissipation of the vortex motion is minimized
when the lattice orientation is aligned parallely with the direction
of its motion \cite{SchmidHauger}. The parallel orientation has been
supported by recent theoretical and numerical simulation studies
\cite{MovingGrass1998,Balents,OlsonPRL} including the unique concept
of the shaking temperature quantifying the effective influence of
inhomogenous environment on moving vortices
\cite{KoshelevePRL1994,KoltonPRL1996,Scheidl}. However, the other
lattice orientation, the perpendicular orientation, has been shown
in limited theoretical studies with different approaches
\cite{ValenzuelaPRL2002,NakaiPhysicaC2009,LiPRB2004}. Therefore, it
needs a physical picture which describes fully the puzzling
mechanism for those two orientations, including the reorientation of
moving VLs.

The issue of the orientation of moving VLs would be interesting when
the magnetic field is rotated from normal to the sample surface.
Since the orientation of moving lattices should be aligned parallely
with the tilt direction \cite{KoganPRB1995}, one expects that the
tilt-induced mechanism competes with the flow-induced one when tilt-
and flow-induced orientations are different. In addition, for thin
superconductors, the field inclination leads to the deformation of
VLs in the sample surface frame \cite{Brandt1993}. Our principle
interest is to what extent the lattice deformation (the anisotropy
in VL parameters) could bear upon the reorientation of moving
lattices. To address the issue, in this study, we present the flow-
and tilt-induced (re)orientation of (deformed) moving vortex
lattices observed on weak pinning, amorphous NbGe superconducting
thin films by means of a mode-locking (ML) technique.

This manuscript is organized as follows: In Sec. 2 we describe
sample preparation and characterizations, followed by our
experimental setup for ML measurements. In Sec. 3 we present the
experimental results of ML features and discuss the lattice
orientation and the anisotropy in VL parameters obtained in
different tilt directions. Derived consequences for dynamic phase
diagrams of the lattice orientation are discussed and they are
compared with an extended anisotropic bond-fluctuation theory
\cite{Scheidl,CommentSVtheory}. Finally, we summarize our findings
in Sec. 4.

\section*{2. Experimental}
 Amorphous Nb$_{x}$Ge$_{1-x}$ ($x\approx
78\%$) superconducting thin films were deposited upon Si substrates
by rf magnetron sputtering. We used a 3 inch diameter niobium target
(purity 99.95$\%$) with germanium sheets (purity 99.999$\%$) on top
and applied 70 W rf power to the target. The NbGe films with 0.28
$\mu$m thick were structured into bridge patterns by using metal
masks covering Si substrates. We used two bridge films of which
geometry and material parameters are as follows: For the film 1 the
width $w$ and length $l$ between voltage contacts are 0.10 mm and
4.00 mm, respectively; the superconducting transition temperature
$T_c$ is 4.09 K, the slope $S$ of the second critical field $B_{c2}$
against temperature $T$ near $T_c$ is $\approx$ 2.1 T/K, and the
normal state resistivity $\rho_n$ is 1.44 $\mu\Omega$m. Using the
dirty-limit expression \cite{Kes1983}, the coherence length
$\xi(0)(=1.81\times10^{-8}/\sqrt{ST_c})$ at $T = 0$ K is 6.1 nm, the
penetration depth $\lambda
(0)(=1.05\times10^{-8}\sqrt{\rho_nT_c^{-1}})$ at $T = 0$ K is 0.62
$\mu$m ($ \gg \xi(0)$), and the Ginzburg-Landau parameter $\kappa (=
3.54\times10^{4}\sqrt{S\rho_n})$ is 62. For the film 2 $w = 0.20$
mm, $l = 1.90$ mm, $T_c =$ 3.50 K, $S \approx$ 2.0 T/K, $\rho_n =$
1.46 $\mu\Omega$m, $\xi(0) = 6.9$ nm, $\lambda (0) =$ 0.68 $\mu$m
and $\kappa =$ 68. Because of large $\kappa$, the vortex density $B$
is nearly equal to the applied magnetic field $B_a$ (\emph{i.e.,} $B
\cong B_a$) when the field is applied perpendicularly to the film
surface.

The NbGe thin films have weak pinning properties for vortices in
magnetic fields and temperatures studied. The depinning current
density $J_c$ determined from a current-voltage ($I-V$) curve with 1
$\mu$V criterion is typically $\sim$ 10 - 100 A/cm$^2$ at $T/T_c$=
0.6 and $B/B_{c2}$= 0.4, which is of the order of $J_c$ for weak
pinning, thin crystals of NbSe$_2$ \cite{kkbprl2005}. The pinning
strength can be characterized by the transverse correlation length
$R_c$ being roughly 10 times larger than the lattice spacing
$a_\triangle$ of the regular hexagonal VL, determined from
two-dimensional collective pinning analysis
\cite{Kes1983,LarkinOvchinnikov}. The $I-V$ curve exhibits the
linear flux-flow behavior, of which flow resistivity was well
discussed in Ref. 30.

We used a laboratory-built insert with a single-axis tilt stage. The
film(s) and the tilt stage were immersed into pumped liquid $^4$He
bath, of which vapor pressure was regulated by a throttle valve and
a capacitance manometer. This allows us to precisely control the
temperature of the bath from 4.2 K down to $\sim$2 K.

A sketch in Fig. 1e represents the orientation of the magnetic field
with respect to the film lying on the $xy$ plane (the tilt stage).
The transport current $I$ is applied along the $y$ axis and vortices
flow along the $x$ axis. $\theta$ is the tilt angle of the magnetic
field from normal to the film surface (the $z$ axis), and $\phi$ is
the in-plane angle between the flow direction (the $x$ axis) and the
field projection onto the $xy$ plane.
In our experimental setup, $\theta$ was adjusted by rotating the
tilt stage, and it was measured \emph{in situ} with a Hall probe
attached to the tilt stage (the $xy$ plane). Due to the line tension
of coaxial cables fixed onto the tilt stage, $\theta$ was varied
within a limited range of 0$^\circ$ $ \lesssim\theta\lesssim$
70$^\circ$. By contrast, the direction of the tilt plane was
adjusted by rotating the bridge film about the tilt axis of the
stage before cooling, and two directions of $\phi \approx$ 0$^\circ$
and 90$^\circ$ were studied in the present study.

The mode locking is based on dynamic resonance, which occurs when an
internal frequency $f_{int}=qv/a$, characterizing the periodicity
(velocity modulation) of moving lattices in the flow direction, is
related harmonically to the frequency $f$ of an ac current $I_{ac}$
superimposed on top of the transport (dc) current $I$, \emph{i.e.},
$qv/a=pf$, where $v$ is the average velocity and $p$, $q$ are
integers
\cite{Fiory,Martinoli,BhattacharyaCDW,Takayama1989,Maeda2002,Kawaguchi2006,Harada,Reichhardt}.
This technique allows us detect not only the periodic vortex spacing
$a$ along the flow direction, but also the deformation and
orientation of moving VLs \cite{kkbphysica2008}. It has been shown
that the large amplitude ac current ($I_{ac} \gtrsim I$) changes the
orientation of moving lattices \cite{OkumaPRB2011}. In order to
avoid this effect, in this study, we employ the rf impedance
technique which can detect the mode locking with the \emph{small}
amplitude ac current ($I_{ac} \ll I$)
\cite{Fiory,kkbphysica2010,Iac}. For the technique we used an LCR
meter (Agilent 4285A) with dc bias option (4285A -001), allowing rf
impedance measurements up to 30 MHz after careful calibration for
the coaxial cables. All the measurements were performed after
cooling the films in zero magnetic field.

\begin{figure}[t]
\includegraphics[width=20pc]{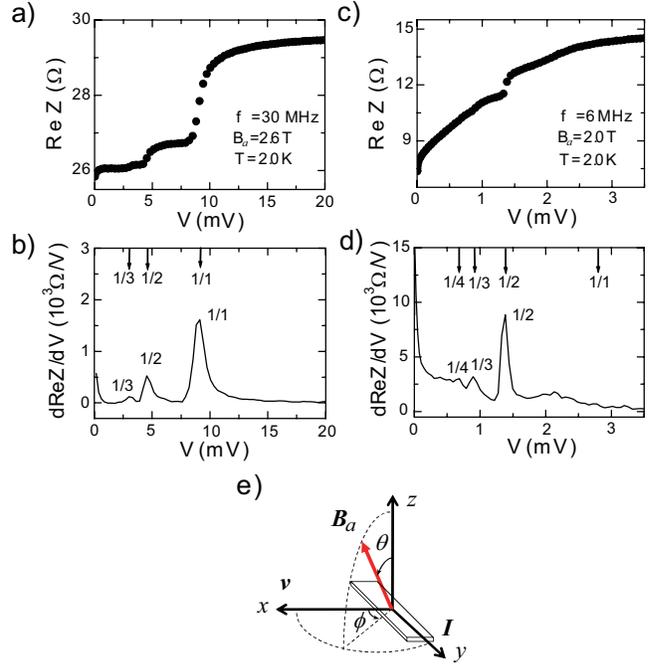}\hspace{2pc}
\caption{(Color online)   Mode-locking features observed in the film
1 at 2.0 K. (a) The real part $Re Z$ of rf impedance vs. voltage $V$
observed at 30 MHz and 2.6 T. The result at 6 MHz and 2.0 T is given
in (c). The derivative curves of (a) and (c) are shown in (b) and
(d), respectively. Arrows in (b) and (d) indicate fundamental and
subharmonic ML voltages. (e) Schematic representation of the
orientation of the applied magnetic field $B_a$ with respect to the
film (the $xy$ plane). The directions of the vortex flow $v$ and the
transport current $I$ are indicated. $\theta$ defines the tilt angle
of the applied magnetic field $B_a$ with respect to normal to the
sample surface (the $z$ axis). The tilt plane in which $B_a$ rotates
is characterized by the in-plain angle $\phi$ between the flow
direction and the field projection on to the film surface.}
\end{figure}

\section*{3.  Results and Discussion}
\subsection*{3.1 Moving vortex lattices in a perpendicular magnetic field}
To provide the physical picture(s) of the mode locking for moving
VLs, let us show some of ML results observed in the flux-flow state
for the magnetic field applied perpendicularly to the sample surface
($\theta = 0^\circ$). During the ML experiments, we ramped up/down
dc current, with superimposing the small, constant amplitude of ac
current on top \cite{Iac}: As responses, we measured both the
flux-flow (dc) voltage $V$ and the real part $ReZ$ of the complex rf
impedance. To display how the complex impedance shows a ML
feature(s), we plot $ReZ$ against $V$. The plot thus obtained in the
film 1 is shown in Fig. 1a. One can see that $ReZ$ exhibits multiple
jumps, as observed in previous studies \cite{Fiory,kkbphysica2010}.
The ML feature can be more clearly displayed by plotting the
derivative of the $ReZ-V$ curve in Fig. 1b, where multiple peaks of
d$ReZ/$d$V$ are visible. These can be identified with the
fundamental ($p/q =$ 1/1) and subharmonics ($p=$1, $q>$1) using the
following arguments: the d$ReZ/$d$V-V$ curve has three peaks with
different height. Focusing on the corresponding voltages for the
peaks, one can find that the voltage for the largest peak is nearly
two (three) times larger than one for the second largest peak (the
smallest peak). Consequently, we find that the largest peak
corresponds to fundamental, and the second and third ones are
subharmonics of $p/q =$ 1/2 and 1/3, respectively. No harmonic with
$p \geq$ 2 appears because of the small amplitude ac current
\cite{kkbprl2002,kkbprb2004}. Another $ReZ-V$ curve is shown in Fig.
1c. The differential plot of the $ReZ-V$ curve (Fig. 1d) shows
clearly that the largest and second largest peaks correspond to 1/2
and 1/3 subharmonics, respectively. The smallest peak is 1/4
subharmonic, and the fundamental peak is not visible. Those two sets
of the ML features with the fundamental largest peak (FLP) and 1/2
subharmonic largest peak (SLP) are typical for the present study.

The observed features originate from the hexagonal orientation of
moving VLs. In previous studies made on amorphous MoGe films
\cite{kkbphysica2008,kkbphysica2010,OkumaPRB2009}, we have shown
that the lattice orientation is characterized by whether one of
close-packed (CP) directions of the hexagonal VL is parallel or
perpendicular to the flow direction. For the parallel orientation,
as depicted in Fig. 1a in Ref. 12, the periodic lattice spacing $a$
in the flow direction is identified with the VL spacing,
\emph{i.e.}, $a=a_\triangle$, indicating the internal frequency of
the moving lattice to be $f_{int}=qv/a_\triangle$. Since the vortex
motion induces a voltage $V=lvB$ over the bridge film with length
$l$, the voltage condition for the mode locking is given by
\begin{equation}
V_{p/q}^\parallel = \frac{p}{q}lfB_aa_\triangle \label{Vparallel}.
\end{equation}
We note that, in the present study, the harmonic number is always
unity ($p=1$) because of the small ac current \cite {kkbprb2004}.
Assuming the lattice spacing for the regular hexagonal VL,
\emph{i.e.}, $a_\triangle=\sqrt{(2/\sqrt{3})\Phi_0/B_a}$ with a
magnetic flux quantum $\Phi_0 (=h/2e)$, we estimate the fundamental
and subharmonic ML voltages and denote them by arrows in Fig. 1b.
One can see that the arrows are quantitatively in good agreement
with the peaks, including the identification of harmonic $p$ and
subharmonic numbers $q$. Thus, from the agreements we are convinced
that the FLP feature of Figs. 1a and 1b corresponds to the parallel
orientation of moving lattices. The similar argument identifies the
SLP feature of Figs. 1c and 1d with the perpendicular orientation of
moving lattices: As sketched in Fig. 1b in Ref. 12, the
perpendicular orientation is characterized by the periodicity of the
row spacing $a_\perp[=(\sqrt{3}/2)a_\triangle]$ along the flow
direction. Thus, the ML voltage for the perpendicular orientation is
given by
\begin{equation}
V_{p/q}^\perp = \frac{p}{q}lfB_a2a_\perp \label{Vperpendicular}.
\end{equation}
Those multiple subharmonics and fundamental are important for
identifying the lattice orientation, however the conclusion does not
change as long as one of identified ML features is given. To make
the argument simple, we focus on the ML voltage at the largest peak
of the FLP (SLP) feature.

The lattice orientation observed above depends on the vortex
velocity and a flow-induced reorientation occurs as the vortex
velocity (frequency) increases. This behavior can be clearly seen by
showing how the ML feature changes with frequency. Figure 2a shows
the results obtained in the film 2 at $B_a=$ 0.6 T and $T$ = 2.5 K.
As depicted by open and solid square symbols, the SLP feature
indicative of the perpendicular orientation appears in low
frequencies, while the FLP one of the parallel orientation in high
frequencies. Thus, from the ML features, one can immediately find
that the flow-induced reorientation occurs at $\approx$ 11 MHz.

The same conclusion can be drawn from the magnitude of the ML
voltage. As observed, for both features the corresponding voltage
increases linearly with frequency, but there is difference in the
slope related to the VL parameters. A solid line represents the ML
voltage condition for the parallel orientation [Eq. (1) with
$p/q=1/1$], while a dotted line does for the perpendicular
orientation [Eq. (2) with $p/q=1/2$]. These are obtained simply by
substituting the sample length $l$ and the magnetic field strength
$B_a$ into Eqs. (1) and (2), without adjustable parameters. One can
see that the former line agrees quantitatively with the higher
frequency results and the latter line does the lower frequency ones.
Thus, the reorientation of moving lattices is marked by a small jump
of the ML voltage at the reorientation frequency $f_r \approx$ 11
MHz.

Figure 2b shows the corresponding $I-V$ curve. It exhibits the
linear flux-flow behavior and does not accompany any noticeable
anomaly, indicating that no discontinuous change in the vortex
velocity or the vortex density occurs at the reorientation. The
corresponding reorientation velocity is $v_r=f_ra_\triangle \approx
0.6 $ m/s, which is two decades smaller than the critical velocity
$v_c \sim 100$ m/s for the flux-flow instability
\cite{Kunchurprb2002}. These results indicate that a proposed
(reorientation) mechanism based on the relaxation of the order
parameter \cite{Kunchurprb2002,Vodolazov2007,OkumaPRB2011} is not
relevant to the observed reorientation.

Additional measurements of the reorientation made in different
magnetic fields (vortex density) provide some insight into the
mechanism for the reorientation. As plotted in Fig. 2c the
reorientation frequency increases with the field and takes a maximum
at $B_a/B_{c2} \approx 0.3$. After that, it decreases with
increasing the field. The broad peak behavior of $f_r$ is
qualitatively in agreement with the field dependence of the shear
elastic modulus $c_{66}$ of the regular VL represented by a solid
curve \cite{Berghuis,Brandt1986}. This qualitative coincidence
suggests that the elastic response of moving lattices plays an
important role upon the flow-induced reorientation.

\begin{figure}[t]
\includegraphics[width=20pc]{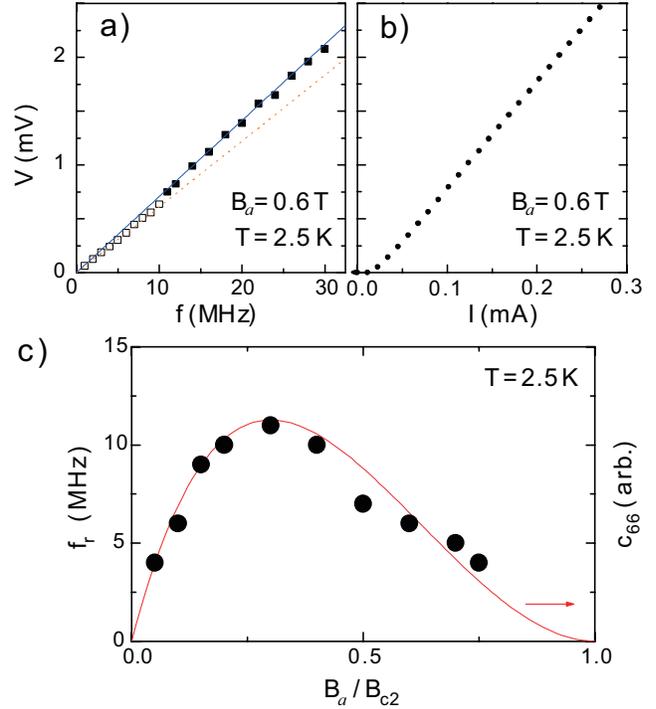}\hspace{2pc}
\caption{(Color online)  (a) ML voltage vs. frequency observed in
the film 2 at 2.5 K and 0.60 T. Solid and dotted lines represent the
ML conditions for parallel and perpendicular orientations,
respectively. (b) The corresponding $I-V$ characteristic measured
simultaneously. (c) Reorientation frequency $f_r$ plotted against
the magnetic field normalized by the second critical field
$B_{c2}($2.5 K$)= 2.0$ T. A solid curve represents the field
dependence of the shear elastic modulus $c_{66}$ of the regular VL.}
\end{figure}

\subsection*{3.2 Moving vortex lattices in an inclined magnetic field}

To track further the physical origin for the reorientation, we study
how the field rotation distorts the moving lattice and also how this
influences on the two orientations induced by the vortex flow. Since
the London theory implies that the orientation of deformed lattices
should be aligned parallely with the tilt direction, the
tilt-induced mechanism competes with the flow-induced one when the
tilt- and flow- induced orientations are different. In this section,
we focus on the ML results observed in two tilt directions of $\phi
= 0^\circ$ and $90^\circ$, and present how the flow-induced
orientation changes into the tilt-induced one as function of tilt
angle $\theta$.

Figure 3a shows the tilt dependence of the ML voltage observed in
the tilt direction of $\phi \approx 90^\circ$. As denoted by open
symbols, the SLP feature, indicative of the perpendicular
orientation, is observed for all the tilt angles studied. This is
consistent with the tilt-induced picture where one of CP directions
of the hexagonal VL should be aligned parallely with the tilt
direction \cite{KoganPRB1995}. The corresponding ML voltage
decreases with increasing $\theta$, and this reduction is consistent
with the dotted curve representing the ML voltage condition for
$\phi \approx 90^\circ$:
\begin{equation}
V_{1/2}^\perp (\theta) = lfB_aa_\perp\cos\theta.
\end{equation}
Here, we assume that the vortex density varies as
$B(\theta)=B_a\cos\theta$ and the lattice parameter along the flow
direction remains unchanged (Consequently, the lattice parameter
along the tilt direction is stretched by $1/\cos\theta$ as
illustrated in the inset to Fig. 3a). We note that the results were
taken at $f=$ 6 MHz ($< f_r$) and the flow-induced orientation (at
$\theta \approx 0 ^\circ$) is the perpendicular one. Thus, the flow-
and tilt-induced orientation coincide, and the perpendicular
orientation appears irrespective of $\theta$.

The situation is different at high frequencies $f > f_r$. The
flow-induced, parallel orientation appears at $\theta \approx 0
^\circ$ and it should switch to the tilt-induced, perpendicular
orientation by rotating the field. This is clearly observed in Fig.
3b, where the ML results taken at 30 MHz ($> f_r$) are plotted. As
observed, the FLP feature (the solid symbol) changes into the SLP
one (the open symbol) as $\theta$ increases, and it occurs at
$\theta \approx 43^\circ$. In other words, the parallel orientation
persists up to $\theta \approx 43^\circ$. This is against the above
picture based on the London theory that the lattice orientation
should be aligned parallely with the tilt direction at any non-zero
tilt angle $\theta \neq 0^\circ$ \cite{KoganPRB1995}.  The
corresponding ML voltages reveal how the moving lattices deform with
increasing $\theta$. The ML condition for the parallel orientation
is given as
\begin{equation}
V_{1/1}^\parallel (\theta) = lfB_a a_\triangle\cos\theta.
\end{equation}
This is simply $2/\sqrt{3}$ times larger than the perpendicular
condition $V_{1/2}^\perp (\theta)$ in Eq. (3) since
$a_\triangle=(2/\sqrt{3})a_\perp$ for the regular VL. As shown in
Fig. 3b, the solid curve representing the parallel condition
explains well the results until a step like drop at the
reorientation angle. This quantitative agreement evidences that the
moving lattices are expanded along the tilt direction. Thus, in the
range of $0^\circ < \theta \leq 43^\circ$ the deformation and
orientation of moving lattices do not coincide in direction as
illustrated in the inset to Fig. 3b.

The discrepancy between the deformation and orientation of moving
lattices also occurs in the other tilt direction of the magnetic
field, \emph{i.e.}, $\phi = 0^\circ$. Since the flow direction
coincides with the tilt direction, one can expect that the
deformation and orientation of moving lattices are aligned parallely
with the flow (and tilt) direction as illustrated in the inset to
Fig. 3d. However, the ML features taken at the low frequency (6 MHz)
reveal the presence of the perpendicular orientation at nonzero tilt
angles. As shown in Fig. 3c, the perpendicular orientation persists
up to $\theta \approx 30^\circ$, above which the parallel
orientation appears. Thus, in the range of $0^\circ < \theta \leq
30^\circ$ the orientation of moving lattices is not aligned
parallely with the flow (and tilt) direction. The lattice
deformation can be found from the tilt dependence of the
corresponding ML voltage. As represented by solid and dotted lines
in Fig. 3c, for both orientations the ML voltage condition is
independent of $\theta$, since the expansion of the lattice
parameter along the flow (and tilt) direction by $1/\cos\theta$ (see
the inset) is canceled with the reduction of the vortex density
$\propto \cos\theta$. As observed, the results follow nicely the
perpendicular condition (the dotted line) until $\theta = 30^\circ$,
above which they do the parallel condition (the solid line). Thus,
the moving lattices deform in the flow (and tilt) direction indeed,
but their orientation is not aligned parallely with the direction at
low tilt angles (see the inset). Since the perpendicular orientation
appears at $\theta \approx 0^\circ$, the discrepancy originates from
the fact that the flow-induced orientation, being different from the
tilt-induced one, survives at low tilt angles. This is not
noticeable in the high frequency (30 MHz) results given in Fig. 3d
where the parallel orientation appears for all the tilt angles,
since the flow-induced orientation coincides with the tilt-induced
one.

\begin{figure}[t]
\includegraphics[width=20pc]{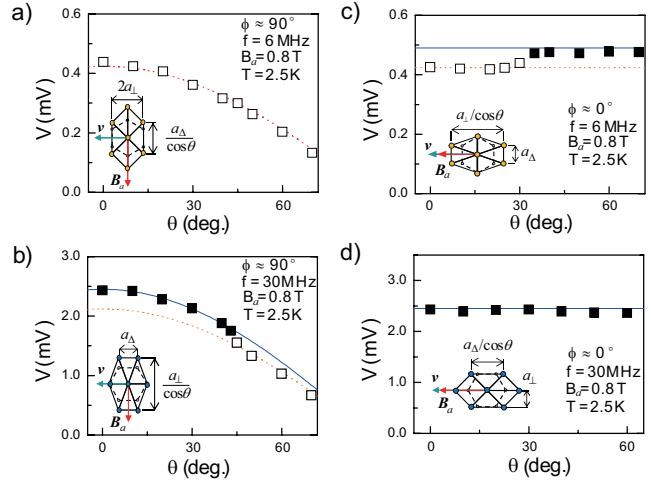}
\caption{(Color online)  ML voltage vs. tilt angle observed in the
film 2 at 2.5 K and 0.80 T.  Deformed moving VLs are schematically
illustrated. The flow and tilt directions are indicated by blue and
red arrows, respectively.}
\end{figure}



Let us discuss the anisotropy of VL parameters with respect to the
flow direction: We define the anisotropy $\gamma$ as the ratio of
the lattice parameter along the flow direction divided by one
perpendicular to the flow direction. Taking the perpendicular
orientation of moving lattices, for instance, we obtain
$\gamma^\perp(\theta)\equiv a_\perp(\theta)/a_\triangle(\theta)
=a^2_\perp(\theta) B_a\cos\theta/\Phi_0 =
[V_{1/2}^\perp(\theta)]^2/(l^2f^2\Phi_0B_a\cos\theta)$. For the
parallel orientation it needs simply to replace
$V_{1/2}^\perp(\theta)$ with $V_{1/1}^\parallel(\theta)$,
\emph{i.e.}, $\gamma^\parallel(\theta)\equiv
a_\triangle(\theta)/a_\perp(\theta)
=[V_{1/1}^\parallel(\theta)]^2/(l^2f^2\Phi_0B_a\cos\theta)$. The
anisotropy thus obtained from the results in Figs. 3a ($\phi \approx
90^\circ$ and $f =$ 6 MHz) and 3d ($\phi \approx 0^\circ$ and $f =$
30 MHz) is summarized in Fig. 4. For clarity we normalize the
results by the ratio of the VL parameters of the regular VL at
$\theta = 0 ^\circ$, \emph{i.e.},
$\overline{\gamma}^\perp(\theta)=\gamma^\perp(\theta)/(\sqrt{3}/2)$
and $\overline{\gamma}^\parallel(\theta) =
\gamma^\parallel(\theta)/(2/\sqrt{3})$. One finds that
$\overline{\gamma}^\perp(\theta)$ varies approximately as
$\cos\theta$ represented by a dotted curve. This agrees with the
picture that the moving lattices are stretched perpendicularly to
the flow direction as discussed above (see the inset to Fig. 3a).
Meanwhile, $\overline{\gamma}^\parallel(\theta)$ increases as
$1/\cos\theta$ (represented by a solid curve) with increasing
$\theta$, being consistent with the picture of the lattice expansion
along the flow direction (see the inset to Fig. 3d). These
agreements give a further support for the description of lattice
deformations illustrated in Fig. 3.

\begin{figure}[t]
\includegraphics[width=15pc]{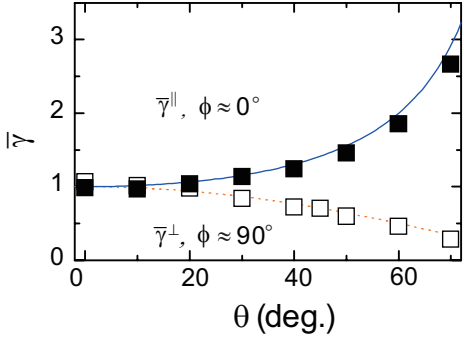}
\caption{(Color online)  Anisotropy in the VL parameters of deformed
moving VLs determined from the ML results in Figs. 3a ($\phi \approx
90^\circ$) and 3d ($\phi \approx 0^\circ$).}
\end{figure}

The above findings related to the lattice orientation imply that the
tilt-induced reorientation occurs, but it depends on not only the
tilt direction $\phi$, but also the vortex velocity (frequency). To
give further insight into the reorientation of moving lattices, it
is useful to trace how the reorientation frequency $f_r$ varies with
$\theta$. To obtain this trace, for each tilt angle $\theta$, we
measured the frequency dependence of the mode locking, and
determined the reorientation frequency $f_r(\theta)$ (see Fig. 2a).
The results thus obtained in the tilt direction of $\phi \approx
90^\circ$ are given in Fig. 5a. As the field is rotated, the
reorientation frequency exhibits a rapid increase with downward
curvature and it reaches the maximum frequency of 30 MHz at $\theta
\approx 43^\circ$. Since $f_r (\theta)$ separates the parallel
orientation from the perpendicular orientation of moving lattices,
the results can be viewed as the dynamic phase diagram for the
lattice orientation: Namely, the region of the parallel orientation
lies above the $f_r (\theta)$ curve, while that of the perpendicular
orientation appears below the curve. In other words, for high tilt
angles $(> 43^\circ)$, the tilt-induced perpendicular orientation
appears irrespective of the vortex velocity. For low tilt angles $(<
43^\circ)$ the lattice orientation depends on the vortex velocity
and the flow-induced reorientation occurs. The corresponding
reorientation velocity increases as the lattice deformation becomes
larger (the anisotropy $\overline{\gamma}$ becomes smaller).
Different results are obtained in the tilt direction of $\phi
\approx 0^\circ$ where the moving lattices are stretched in the flow
direction. As shown in Fig. 5b, $f_r (\theta)$ decreases with
increasing $\theta$ and it seems to vanish at $\theta \approx
40^\circ$. Namely, for low tilt angles $(< 40^\circ)$ the more
stretched the moving lattices are, the smaller the reorientation
frequency becomes. As a result, the region of the perpendicular
orientation shrinks and the tilt-induced parallel orientation
dominates high tilt angles $(> 40^\circ)$. From these results we
remark that (1) for high tilt angles ($\gtrsim 40^\circ$) the field
inclination determines the orientation of moving VLs: (2) For low
tilt angles ($\lesssim 40^\circ$) the vortex velocity determines the
lattice orientation: (3) The corresponding reorientation velocity
varies sensitively as the moving lattices deform and it increases
(decreases) as the anisotropy $\overline{\gamma}$ becomes small
(large).

\begin{figure}[t]
\includegraphics[width=17pc]{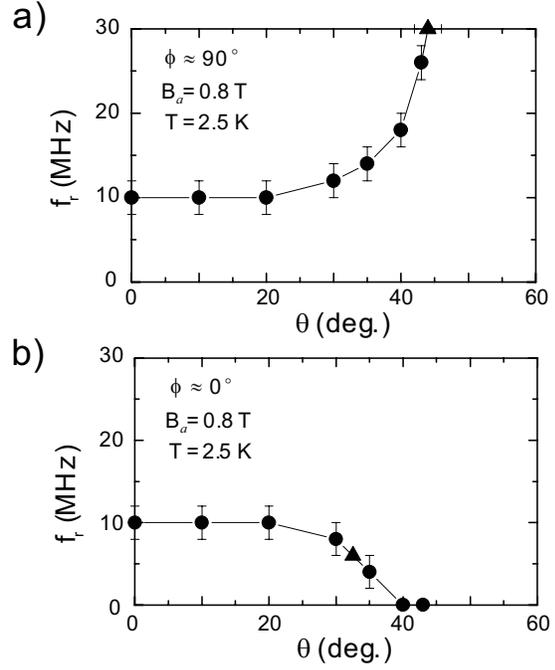}\hspace{2pc}
\caption{Reorientation frequency vs. tilt angle (a) for $\phi
\approx 90^\circ$ and (b) for $\phi \approx 0^\circ$ at 2.5 K and
0.8 T. The triangle symbols indicate the reorientation angles
determined from the tilt dependences of the ML voltage given in
Figs. 3b and 3c. Lines are guide to the eye.}
\end{figure}


The argument of the London theory indicates that the orientation of
the vortex lattice should be aligned parallelly with the tilt
direction as soon as the magnetic field is tilted from normal to the
film surface \cite{KoganPRB1995}. This picture disagrees with our
experimental findings for low tilt angles ($\lesssim 40^\circ$).
Therefore a factor(s) other than the vortex-vortex interaction
should be relevant to the observed lattice orientations. Since the
vortices are driven over the disordered (pinning) environment, let
us discuss the above results from the viewpoint of the anisotropic
shaking motion of moving lattices. Scheidl and Vinokur (SV) have
introduced pinning-bond widths characterizing the relative
displacements of two neighboring vortices and have argued that the
anisotropy in the bond widths could determine the orientation of
moving lattices \cite{Scheidl}. When the bonds along the flow
direction ($x$-bonds) fluctuate weakly than those perpendicular to
the flow ($y$-bonds), the moving vortices are aligned parallelly
with the flow direction. This results in the parallel orientation of
moving lattices. In opposite case, the vortices should be aligned
perpendicularly to the flow direction, resulting in the
perpendicular orientation of moving lattices \cite{CommentSVtheory}.
The dependences of $x$- ($w_x^{pin}$) and $y$-bond widths
($w_y^{pin}$) on the vortex velocity $v$ are given in limited cases
\cite{Scheidl}: For small velocities the elastic interaction
dominates and both bond widths diverge as $\sim \ln(c_{66}/v)$ at $v
\rightarrow$ 0. However, prefactors show that the $x$-bond width is
larger than the $y$-bond width. Thus their ratio $\gamma_{w}(\equiv
w_x^{pin}/w_y^{pin})$ is more than 1 and the perpendicular
orientation is favorable at small velocities \cite{CommentSVtheory}.
As the velocity increases, the width of $x$ bonds decreases as
$1/v^2$, which is faster than the $y$ bond width $\propto 1/v$,
indicating the parallel orientation ($\gamma_{w} <$ 1) at large
velocities. Thus, the reorientation marks the point where the widths
of $x$ and $y$ bonds become comparable, \emph{i.e.}, $\gamma_{w}
\approx$ 1. We expect that this condition depends how the moving
lattices deform and is scaled with the anisotropy of the lattice
parameters of moving lattices, \emph{i.e.}, $\gamma_w \approx
\overline{\gamma}$. For $\phi = 90^\circ$ the condition gives
$\gamma_w \approx \cos \theta (< 1)$. Namely, the $x$-bond width is
smaller than the $y$-bond width, and this requires that the
reorientation velocity is faster than one for no deformation at the
perpendicular field ($\theta = 0^\circ$). This is consistent with
our observation that $f_r$ increases with $\theta$ (see Fig. 5a).
For $\phi = 0^\circ$ the condition reads $\gamma_w \approx 1/ \cos
\theta (> 1)$. Namely, the $x$-bond width is larger than the
$y$-bond width, and thus the reorientation velocity becomes slower.
This explains qualitatively the reduction of $f_r$ as $\theta$
increases (see Fig. 5b). A quantitative comparison can be made at
the small velocity limit. According to the SV theory, the anisotropy
of the bond widths is given as $\gamma_{w}= 1+(\sqrt{2\pi
B_a/\Phi_0}\xi)/(1+\sqrt{0.5 \pi B_a/\Phi_0}\xi)$. Using $B_a=$ 0.80
T and the coherence length $\xi(0) =$ 6.9 nm for the film 2, we
obtain $\gamma_{w} =$ 1.3. This is quantitatively consistent with
the lattice expansion $\overline{\gamma} = 1/\cos\theta = 1.3$ at
$\theta \approx 40^\circ$ at which $f_r$ vanishes ($v \approx$ 0).
These agreements suggest strongly that the anisotropic shaking
motion plays an essential role for determining the orientation of
moving VLs.

A further test of the SV model can be made on confined vortex
systems such as narrow constrictions
\cite{kkbprl2002,Jukna2002,Vlasko-Vlasov2002,Silhanek2009,Yu2010,Wordenweber2012}
or possibly small disks \cite{Lopez1999,CorbinoSimulation}. The
shear rigidity of moving lattices appears in the transport
properties such as the flux-flow resistance, and the anisotropy in
VL parameters can be tuned by the matching condition between the
vortex density and the size of the constriction. These would allow
systematic and quantitative investigation on how the reorientation
frequency varies with the anisotropy of VL parameters in confined
moving lattices.

\section*{4. Summary}
In summary, we have presented ML experiments of moving VLs in
amorphous NbGe superconducting films. The rf ML technique allows us
to find the orientation of moving lattices as function of frequency
(the vortex velocity). The flow-induced reorientation of moving
lattices occurs at the characteristic frequency $f_r$, separating
the perpendicular orientation at small velocities from the parallel
orientation at large velocities. The magnetic field dependence of
$f_r$ resembles qualitatively with that of the shear elastic modulus
$c_{66}$ of the regular VL, suggesting the relevance of the
elasticity of moving lattices on the reorientation.

Introducing the deformation of moving lattices by rotating the field
in different directions, we have shown how the lattice orientation
varies with tilt angle. The moving lattice is expanded in the
direction of its motion when the field is rotated along the flow
direction ($\phi \approx 0^\circ$), while for the field rotated
perpendicularly to the flow direction ($\phi \approx 90^\circ$) the
moving lattices are stretched perpendicularly to the flow. For high
tilt angles the orientation of moving lattices is aligned parallely
with the tilt direction, being consistent with the London theory.
Meanwhile for low tilt angles the lattice orientation depends on the
vortex velocity and the flow-induced reorientation occurs. Focusing
on the results for low tilt angles, we have traced how the
reorientation frequency $f_r$ varies with tilt angle. The
flow-induced reorientation needs large moving velocities when the
moving lattices are stretched perpendicularly to the flow direction,
while it occurs at small velocities when the lattices are expanded
in the flow direction. Thus, the anisotropy in VL parameters is
crucial for the flow-induced reorientation. The observed features
are consistent with the extended picture of the anisotropic bond
fluctuation theory. This reveals that the anisotropic shaking motion
is the essential mechanism which governs the orientation of moving
VLs.

\begin{acknowledgment}


N. K. thanks P. H. Kes, V. M. Vinokur, M. R. Eskildsen, T. Nishizaki
and S. Okuma for stimulating discussions. This work was supported
partly by the grant in Aid for Scientific research from MEXT (the
Ministry of Education, Culture, Sports, Science and Technology),
Japan.

\end{acknowledgment}


\begin{thebibliography}{9}
\bibitem{CommentSVtheory} We note that the possibility of the perpendicular
orientation at small velocities is not claimed in Ref. 21, although
the analytical results of bond widths at the small velocity limit
($x$-bond width is larger than $y$-bond width) are derived.
\bibitem{Iac} The magnitude of the superimposed ac current was 7 $\mu$A ($< I_c \approx$ 20 $\mu$A). Meanwhile, the dc
current was ramped up to $\sim$ 1 mA in order to observe the ML
feature in the flux-flow state.
\end{thebibliography}
\end{document}